\documentclass[preprint,showpacs,preprintnumbers,amssymb,amsmath]{revtex4}

\usepackage{graphicx}% Include Fig. files
\usepackage{dcolumn}% Align table columns on decimal point
\usepackage{bm}% bold math

\begin{document}

\title{Study of the  $D^0 \rightarrow \pi^-\pi^+\pi^-\pi^+$ decay.}

\affiliation{University of California, Davis, CA 95616}
\affiliation{Centro Brasileiro de Pesquisas F\'\i sicas, Rio de Janeiro, RJ, Brasil}
\affiliation{CINVESTAV, 07000 M\'exico City, DF, Mexico}
\affiliation{University of Colorado, Boulder, CO 80309}
\affiliation{Fermi National Accelerator Laboratory, Batavia, IL 60510}
\affiliation{Laboratori Nazionali di Frascati dell'INFN, Frascati, Italy I-00044}
\affiliation{University of Guanajuato, 37150 Leon, Guanajuato, Mexico}
\affiliation{University of Illinois, Urbana-Champaign, IL 61801}
\affiliation{Indiana University, Bloomington, IN 47405}
\affiliation{Korea University, Seoul, Korea 136-701}
\affiliation{Kyungpook National University, Taegu, Korea 702-701}
\affiliation{INFN and University of Milano, Milano, Italy}
\affiliation{University of North Carolina, Asheville, NC 28804}
\affiliation{Dipartimento di Fisica Nucleare e Teorica and INFN, Pavia, Italy}
\affiliation{Pontif\'\i cia Universidade Cat\'olica, Rio de Janeiro, RJ, Brazil}
\affiliation{University of Puerto Rico, Mayaguez, PR 00681}
\affiliation{University of South Carolina, Columbia, SC 29208}
\affiliation{University of Tennessee, Knoxville, TN 37996}
\affiliation{Vanderbilt University, Nashville, TN 37235}
\affiliation{University of Wisconsin, Madison, WI 53706}
\author{J.~M.~Link}
\affiliation{University of California, Davis, CA 95616}
\author{P.~M.~Yager}
\affiliation{University of California, Davis, CA 95616}
\author{J.~C.~Anjos}
\affiliation{Centro Brasileiro de Pesquisas F\'\i sicas, Rio de Janeiro, RJ, Brazil}
\author{I.~Bediaga}
\affiliation{Centro Brasileiro de Pesquisas F\'\i sicas, Rio de Janeiro, RJ, Brazil}
\author{C.~Castromonte}
\affiliation{Centro Brasileiro de Pesquisas F\'\i sicas, Rio de Janeiro, RJ, Brazil}
\author{A.~A.~Machado}
\affiliation{Centro Brasileiro de Pesquisas F\'\i sicas, Rio de Janeiro, RJ, Brazil}
\author{J.~Magnin}
\affiliation{Centro Brasileiro de Pesquisas F\'\i sicas, Rio de Janeiro, RJ, Brazil}
\author{A.~Massafferri}
\affiliation{Centro Brasileiro de Pesquisas F\'\i sicas, Rio de Janeiro, RJ, Brazil}
\author{J.~M.~de~Miranda}
\affiliation{Centro Brasileiro de Pesquisas F\'\i sicas, Rio de Janeiro, RJ, Brazil}
\author{I.~M.~Pepe}
\affiliation{Centro Brasileiro de Pesquisas F\'\i sicas, Rio de Janeiro, RJ, Brazil}
% ...
\collaboration{The FOCUS Collaboration}
\affiliation{See \textrm{http://www-focus.fnal.gov/authors.html} for additional author information.}

\date{\today}% It is always \today, today,

\begin{abstract}
Using data from the FOCUS (E831) experiment at Fermilab, we present new measurements for 
the Cabibbo-suppressed decay mode $D^0 \rightarrow \pi^-\pi^+\pi^-\pi^+$. We measure the branching ratio
$\Gamma(D^0 \to\pi^+\pi^- \pi^+\pi^-)/\Gamma(D^0 \to K^-\pi^+\pi^-\pi^+) = 0.0914 \pm 0.0018 \pm 0.0022$.
An amplitude analysis has been performed, a first for this channel, in order to determine the resonant 
substructure of this decay mode. 
The dominant component is the decay $D^0 \to a_1(1260)^+ \pi^-$, accounting for 60\% of the decay rate.
The second most dominant contribution comes from the decay $D^0 \to \rho(770)^0\rho(770)^0$,
with a fraction of 25\%. We also study the $a_1(1260)$ line shape and resonant substructure.
Using the helicity formalism for the angular distribution of the decay $D^0 \to \rho(770)^0\rho(770)^0$,
we measure a longitudinal polarization of $P_L = (71 \pm 4\pm 2)$\%.

\end{abstract}

\maketitle

%%%%%%%%%%%%%%%%%%%%%%%%%%%%%%%%%%%%%%%%%%%%%%%%%%%%%%%%%%%%%%%%%%%%%%%%%%%%%%%%
\section{\label{sec:level1}Introduction}
%%%%%%%%%%%%%%%%%%%%%%%%%%%%%%%%%%%%%%%%%%%%%%%%%%%%%%%%%%%%%%%%%%%%%%%%%%%%%%%%

Hadronic decays of charm mesons are an important tool for understanding the dynamics of the
strong interaction in the low energy regime. Hadronic decays of $D$ mesons typically have a rich 
resonant structure and 
a small nonresonant component. Scalar mesons are abundant products of three-body decays which have a pair of 
identical pions in the final state. Due to these features, three-body decays of $D$ mesons have been extensively used
to study the $\pi\pi$ and $K\pi$ systems, with emphasis on the S-wave component of their amplitudes 
and on the scalar resonances \cite{e791,malvezzi,bm,cleo1,jim}. With $D$ decays one can continuously cover the 
$\pi\pi$ and $K\pi$  mass spectra from threshold up to 1.7 $\mathrm{GeV}/c^2$ ($M_D-m_{\pi}$), filling the 
existing gaps between the $K_{e4}$ \cite{kl4} and the CERN-M\"{u}nich data \cite{cern-munich}, 
in the $\pi\pi$ case, and between 
threshold and 825 $\mathrm{MeV}/c^2$, where the LASS data \cite{LASS} on $K\pi$  scattering start. Another interesting 
feature of $D$ decays is that the bulk of the hadronic decay width can be described in terms of simple tree-level quark
diagrams. There seems to be, at least for the vector, axial-vector and 
tensor mesons, a strong correlation between the final state quarks --- the spectator valence quark plus the ones
resulting from the weak decay of the $c$ quark --- and the quark content of the observed resonances in the 
intermediate states. This connection allows some insights on the nature of light mesons. The information provided 
by $D$ decays is, therefore, complementary to that of traditional scattering experiments. 

Much less information is available concerning the $\pi\pi\pi$/$K\pi\pi$  systems and the axial-vector 
resonances. The picture emerging from many different Dalitz plot analyses of $D$ decays reveals
a well defined pattern: final states that can be associated with a simple spectator amplitude ($W$-radiation),
in which the virtual $W$ couples to a vector or axial-vector meson, have large branching fractions 
compared to final states in which the $W$ is coupled to a pseudo-scalar meson. Following this pattern, one expects
the $a_1(1260)^+ \pi^-$ channel to be the dominant resonant component of the 
$D^0 \to \pi^-\pi^+\pi^-\pi^+$ decay (charge conjugation is always implied, unless stated otherwise).
This decay, therefore, can be used to study the resonant substructure of
the axial-vector meson $a_1(1260)$, as well as its line shape, from threshold up to 1.72 $\mathrm{GeV}/c^2$.

In the study of light mesons from $D$ decays there is one potential difficulty: in hadronic decays 
the $\pi\pi$ and $K\pi$ pairs,
as well as the $\pi\pi\pi$/$K\pi\pi$  systems, are part of a few-body strongly interacting system. 
In principle one must account for final state interactions (FSI) between all decay particles. 
It is still an open question whether the FSI are strong enough in three-body decays to 
distort the pure $\pi\pi$/$K\pi$ scattering amplitudes. In any case, the effects 
of the FSI should become more important as the number of final state particles increases. 

Most amplitude analyses employ the so called isobar model: a coherent sum of resonant amplitudes weighted by 
constant complex coefficients. The constant phases account for re-scattering effects between the isobar and the 
other decay particles. This approach to FSI may be sufficient in some cases, but may be too simplistic in four-body decays. 
In the most general case, the FSI would depend on energy, with a smooth variation across the phase space.
A correction to the isobar model would then be necessary in order to incorporate the energy dependent FSI.
As we will show, the energy dependent FSI can probably not be ignored in four-body decays of $D$ mesons.

There is an additional motivation for the study of four-body $D$ meson decays. Charmless
decays of $B$ mesons are a promising tool for the study of \emph{CP} violation. In particular,  
the decays $B \to \rho\rho$ have been used to extract the CKM angle $\alpha$ \cite{belle,babar,babarvv,babarvv2}.
The $B \to a_1(1260) \pi$ mode, however, also leads to the same $\pi\pi\pi\pi$ final state.
The $B^0 \to a_1(1260)^- \pi^+$ channel has a branching fraction
which is one order of magnitude larger than that of $B^0 \to \rho^0\rho^0$ \cite{babara1,bellea1}. There is no
measurement of the $B^{\pm} \to a_1(1260)^{\pm} \pi^0$ branching ratio, but since this decay proceeds mainly via the same
tree-level $W$ radiation amplitude as that of the $\rho^{\pm}\rho^0$ channel, one expects these two modes
to have comparable rates. The $B \to a_1(1260) \pi$ decay, thus, accounts for a large fraction of the 
$B \to \pi\pi\pi\pi$ decay, and a full amplitude analysis would be necessary in order to isolate the 
$B \to  \rho\rho$ contribution, as in the case of the $D^0 \to \pi^-\pi^+\pi^-\pi^+$ decay.
All the systematics --- form factors, resonance line shapes, representation of the S-wave components, 
angular distributions, FSI, etc. --- are common to both decays. One can think of the $D^0 \to \pi^-\pi^+\pi^-\pi^+$ 
decay as a prototype of the $B\to \pi\pi\pi\pi$ decay. 
 
In this paper, we present a new measurement of the relative branching ratio (BR)
$\Gamma(D^0 \to \pi^-\pi^+\pi^-\pi^+)/\Gamma(D^0 \to K^-\pi^+\pi^-\pi^+)$ 
using data from the FOCUS experiment. For the first time an amplitude analysis has been 
performed to determine the $D^0 \to \pi^-\pi^+\pi^-\pi^+$ resonant substructure.

%%%%%%%%%%%%%%%%%%%%%%%%%%%%%%%%%%%%%%%%%%%%%%%%%%%%%%%%%%%%%%%%%%%%%%%%%%%%%%%%
\section{\label{sec:level1} The FOCUS experiment}
%%%%%%%%%%%%%%%%%%%%%%%%%%%%%%%%%%%%%%%%%%%%%%%%%%%%%%%%%%%%%%%%%%%%%%%%%%%%%%%%

FOCUS, an upgraded version of E687~\cite{spectro}, is a charm photo-production experiment  which collected 
data during the 1996--97 fixed target run at Fermilab. Electron and positron beams 
(typically with
$300~\textrm{GeV}$ endpoint energy) obtained from the $800~\textrm{GeV}$ 
Tevatron proton beam produce, by means of bremsstrahlung, a photon beam which
interacts with a segmented BeO target~\cite{photon}. 
The mean photon energy for reconstructed charm events is $\sim 180~\textrm{GeV}$. A system 
of three multi-cell threshold \v{C}erenkov
counters performs the charged particle identification, separating kaons from
pions up to a momentum of $60~\textrm{GeV}/c$. Two systems of silicon micro-vertex
detectors are used to track particles: the first system consists of 4 planes
of micro-strips interleaved with the experimental target~\cite{WJohns} and the
second system consists of 12 planes of micro-strips located downstream of the
target. These high resolution detectors allow the identification and separation of charm
primary (production) and secondary (decay) vertices. The charged particle
momentum is determined by measuring the deflections in two magnets of
opposite polarity through five stations of multi-wire proportional chambers.

%%%%%%%%%%%%%%%%%%%%%%%%%%%%%%%%%%%%%%%%%%%%%%%%%%%%%%%%%%%%%%%%%%%%%%%%%%%%%%%%
\section{\label{sec:level1}The $D^0 \to \pi^-\pi^+\pi^-\pi^+$ sample}
%%%%%%%%%%%%%%%%%%%%%%%%%%%%%%%%%%%%%%%%%%%%%%%%%%%%%%%%%%%%%%%%%%%%%%%%%%%%%%%%

The final states are selected using a candidate driven vertex
algorithm~\cite{spectro}. To minimize systematic errors on the measurements of the branching ratio,
we use identical vertex cuts on the signal and normalizing mode.

Secondary vertices are formed from the four
candidate tracks. The momentum of the resultant $D^{0}$ candidate is used as
a \textit{seed} track to intersect other reconstructed tracks and to 
search for a primary vertex. The confidence levels of both vertices are
required to be greater than 1\%.
The mean energy of the $D^0$ candidates in the LAB frame is 85GeV. This is equivalent to
an average decay length of 1cm, allowing a good spatial separation between
the production and decay vertices. Once the position of the production and decay of the $D^0$ candidate is
determined, the distance $L$ between the vertices and its uncertainty $\sigma _{L}$ are computed. 
The ratio $L/\sigma _{L}$ is the most important variable
for separating charm events from non-charm prompt backgrounds. Signal quality is further 
enhanced by cutting on the isolation variables, \emph{Iso1} and \emph{Iso2}. The isolation
variable \emph{Iso1} requires that the tracks forming the $D$ candidate vertex 
have a confidence level smaller than the cut to form a vertex with the tracks from the
primary vertex. The \emph{Iso2} variable requires that all remaining tracks not assigned to the 
primary and secondary vertex have a confidence level smaller than the cut to form a vertex 
with the $D$ candidate daughters. In addition, we require the secondary vertex to lie outside of the segmented targets 
({\em O}ut-{\em o}f-{\em M}aterial cut, or {\em OoM}), in order to reduce 
contamination due to secondary interactions. The {\em OoM} variable is actually the
the distance between the secondary vertex and the edge of the nearest target segment
divided by the uncertainty on the secondary vertex location. We have also applied a cut on 
$\mathrm{vtx}_{\mathrm{score}}$, a 
variable built from the {\em OoM}, $L$\thinspace /\thinspace $\sigma _{L}$ and the
confidence level of the secondary vertex, in order to explore the correlations between
these variables. Each of these variables is normalized to its maximum value, so
$\mathrm{vtx}_{\mathrm{score}}$ ranges from 0 to 1. This cut allows us to further reduce the background without
applying tighter cuts on each of these variables. The set of cuts that yield the $D^0 \to \pi^-\pi^+\pi^-\pi^+$
signal with the best statistical significance is
$L$\thinspace /\thinspace $\sigma _{L}$ $>$ $10$, \emph{Iso1} and \emph{Iso2} $<$ 10\%,
$\mathrm{vtx}_{\mathrm{score}}>0.15$ and $OoM>1$.

The only difference in the selection criteria between the  $D^0 \to \pi^-\pi^+\pi^-\pi^+$ and
$D^{0} \to K^-\pi^+\pi^-\pi^+$ decay modes 
lies in the particle identification cuts. The \v{C}erenkov identification cuts used in
FOCUS are based on likelihood ratios between the various particle
identification hypotheses. These likelihoods are computed for a given track
from the observed firing response (on or off) of all the cells that are
within the track's ($\beta =1$) \v{C}erenkov cone for each of our three 
\v{C}erenkov counters. The product of all firing probabilities for all the cells
within the three \v{C}erenkov cones produces a $\chi ^{2}$-like variable 
$W_{i}=-2\log (\mathrm{Likelihood})$ where $i$ ranges over the electron, pion,
kaon, and proton hypotheses~\cite{cerenkov}. The kaon track is required
to have $\Delta _{K}=W_{\pi }-W_{K}>3$, and all 
pion tracks are required to be separated by less than $5$  from the best
hypothesis, that is $\Delta_{\pi}=W_\mathrm{min}-W_{\pi }>-5$.
 
\begin{figure}
\includegraphics[width=8.5 cm]{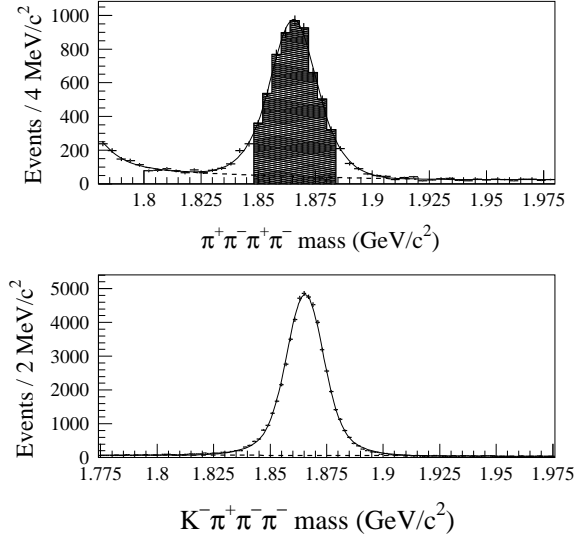}
\caption{Invariant mass distributions used to determine the ratio of branching fractions.
The upper plot is the $\protect\pi^-\protect\pi^+\protect\pi^-\protect\pi^+$ signal;
the lower plot is the normalizing channel $K^-\protect\pi^+\protect\pi^-\protect\pi^+$. 
The fits (solid curves) are explained in the text; 
the dashed line shows the background. The hatched area on the
$\protect\pi^-\protect\pi^+\protect\pi^-\protect\pi^+$ signal
corresponds to the events used in the amplitude analysis.}
\label{m4pi}
\end{figure}

Using the set of selection cuts just described, we obtain the invariant mass distribution 
for $\pi^-\pi^+\pi^-\pi^+$ shown in Fig. \ref{m4pi}. Although the \v{C}erenkov cuts 
considerably reduce  the reflection peak (from $D^0 \to K^-\pi^+\pi^-\pi^+$) to the left of the
signal peak, there is still a large distortion of the background due to this surviving 
contamination. 
The $\pi^-\pi^+\pi^-\pi^+$ mass plot is fit with a function that includes two Gaussians with the 
same mean but different sigmas to take into account the variation in momentum resolution of our 
spectrometer as a function of particle momentum \cite{spectro}, and two exponential functions for the 
combinatorial background and 
for the $D^0 \to K^-\pi^+\pi^-\pi^+$ reflection. A log-likelihood fit gives a signal of 
$6360 \pm 115$ $D^0 \to \pi^-\pi^+\pi^-\pi^+$ events, over a background of $769\pm15$ events.

The large statistics $K^{-}\pi^{-}\pi^{+}\pi^{+}$ mass plot is also fit with two Gaussians with the 
same mean and different sigmas plus a second-order polynomial for the background. 
A log-likelihood fit gives a signal of $54156 \pm 267$  $D^0 \to K^{-}\pi^{-}\pi^{+}\pi^{+}$ events.

%%%%%%%%%%%%%%%%%%%%%%%%%%%%%%%%%%%%%%%%%%%%%%%%%%%%%%%%%%%%%%%%%%%%%%%%%%%%%%%%
%
%                                  BRANCHING RATIO
%
%%%%%%%%%%%%%%%%%%%%%%%%%%%%%%%%%%%%%%%%%%%%%%%%%%%%%%%%%%%%%%%%%%%%%%%%%%%%%%%%

\section{\label{sec:level1}Relative Branching Ratio}

The evaluation of relative branching ratios requires yields from the fits to
be corrected for detector acceptance and efficiency. These differ among the various
decay modes, depending on the $Q$ values and \v{C}erenkov identification efficiency.

From the Monte Carlo simulations, we compute the relative efficiencies (with statistical error only):

\begin{equation} 
\frac{\epsilon(D^0 \to K^-\pi^+\pi^-\pi^+) }{\epsilon(D^{0} \to \pi^-\pi^+\pi^-\pi^+)}
= 0.7891 \pm 0.0004. 
\end{equation}

Using the previous results, we obtain the following value for the branching ratio:

\begin{equation}
\frac{\Gamma(D^0 \to \pi^-\pi^+\pi^-\pi^+)}{\Gamma(D^0 \to K^-\pi^+\pi^-\pi^+)} = 0.0914 \pm 0.0018.
\end{equation}

Systematic uncertainties on the branching ratio measurement may come from
different sources. We estimate
the systematic uncertainty on the yields and on the efficiencies considering three 
independent types of contributions: the \emph{split sample} component, the \emph{fit
variant} component, and the component due to the particular choice of the
vertex and \v{C}erenkov cuts.

The \emph{split sample} component takes into account the systematics
introduced by a residual difference between data and Monte Carlo, due to a
possible mismatch in the reproduction of the $D^{0}$ momentum and the changing
experimental conditions of the spectrometer during data collection. This 
component has been determined by splitting data
into four independent subsamples, according to the $D^{0}$ momentum range
(high and low momentum) and the configuration of the vertex detector,
that is, before and after the insertion of the upstream silicon system. A technique,
employed in FOCUS and modeled after the 
\emph{S-factor method} from the Particle Data Group~\cite{pdg}, was used 
to try to separate true systematic variations from statistical 
fluctuations. The branching ratio is evaluated for each of the four 
statistically independent subsamples and a \emph{scaled error} $\tilde{\sigma}$ (that 
is the errors are boosted when $\chi ^{2}/(N-1)>1$) is calculated.  
The \emph{split sample} error $\sigma_\mathrm{split}$ is defined as the 
difference in quadrature between the statistical error returned by the fit on the unsplit data set,
$\sigma_{stat}$, and the scaled error, if the scaled error exceeds the statistical error, 
$\sigma_\mathrm{split}=\sqrt{\tilde{\sigma}^2-\sigma_{stat}^2}$.

Another possible source of systematic uncertainty is the \emph{fit variant}.
This component is computed by varying, in a reasonable manner, the fitting
conditions for the whole data set. In our study we fixed the widths of the 
Gaussian to the values obtained by the Monte Carlo simulation, we changed 
the background parametrization (varying the degree of the polynomial), and we use one 
Gaussian instead of two. The latter is the dominant contribution to the \emph{fit variant} error.
Finally the variation of the computed efficiencies,  both for 
$D^0 \to \pi^-\pi^+\pi^-\pi^+$ and the normalizing decay mode, due to the different 
resonant substructure simulated in the Monte Carlo has been taken into account. 
The BR values obtained by these variants are all {\em a priori} equally likely, therefore this 
uncertainty can be estimated by the {\it r.m.s.}\ of the measurements.

We estimate the cut component systematic error by varying vertex and particle identification cuts one at
a time. We varied the confidence
level of the secondary vertex from 1\% to 50\%, \emph{Iso1} and \emph{Iso2} from $10^{-6}$ to $1$,
$L$\thinspace /\thinspace $\sigma _{L}$ from $6$ to $20$, $\Delta_{\pi}$ from $-6$ to $-2$, and
$\mathrm{vtx}_{\mathrm{score}}$ from 0 to 0.2.
Analogously to the \emph{fit variant}, the cut component is estimated using 
the standard deviation of the several sets of cuts.
Actually, this is an overestimate of the cut component because the statistics of 
the cut samples are different. 

Finally, adding in quadrature the three components, we get the final systematic 
errors which are summarized in Table~\ref{err_sist}. 

\begin{table}
\caption{Contributions to the systematic uncertainties of 
the branching ratio \\ 
$\Gamma(D^0 \to \pi^-\pi^+\pi^-\pi^+)/\Gamma(D^0 \to K^-\pi^+\pi^-\pi^+)$.}
\label{err_sist}
\begin{ruledtabular}
\begin{tabular}{lcr}
{Source}        & {Systematic error}      \\
\hline
{Split sample}  &  0.0010      \\
{Fit Variant}   &  0.0012      \\
{Set of cuts}   &  0.0016    \\ \hline
{Total systematic error} &  0.0022 \\
\end{tabular}
\end{ruledtabular}
\end{table}

The final result is shown in Table~\ref{comparison} along with a
comparison with previous measurements.

\begin{table}
\caption{Comparison with other experiments.}
\label{comparison}
\begin{ruledtabular}
\begin{tabular}{lcr}
Experiment& $\Gamma(D^0 \to \pi^-\pi^+\pi^-\pi^+)/\Gamma(D^0 \to K^-\pi^+\pi^-\pi^+)$ & Events \\ \hline
FOCUS (this result)& $0.0914  \pm 0.0018 \pm 0.0022 $ & $ 6360 \pm 115 $    \\ 
CLEO-c~\cite{CLEOc}   & $0.097  \pm 0.002  \pm 0.003  $ & $  7331 \pm 130 $     \\
BES~\cite{BES} & $0.079   \pm 0.018  \pm 0.005  $ & $  162 \pm 20 $     \\
E687~\cite{E687}   & $0.095   \pm 0.007  \pm 0.002  $ & $  814 \pm 26 $     \\ 
\end{tabular}
\end{ruledtabular}
\end{table}

%%%%%%%%%%%%%%%%%%%%%%%%%%%%%%%%%%%%%%%%%%%%%%%%%%%%%%%%%%%%%%%%%%%%%%%%%%%%%%%%
%
%                                  Amplitude analysis
%
%%%%%%%%%%%%%%%%%%%%%%%%%%%%%%%%%%%%%%%%%%%%%%%%%%%%%%%%%%%%%%%%%%%%%%%%%%%%%%%%

\section{Amplitude Analysis}

A fully coherent amplitude analysis was performed in order to determine the
resonant substructure of the $D^0 \to \pi^-\pi^+\pi^-\pi^+$ decay. This is, to
our knowledge, the first such measurement for this channel. The amplitude analysis was
performed in the framework of the isobar model on a sample of 6153 events, corresponding to
the hatched area in Fig.~\ref{m4pi} (events with $m_{4\pi}$ within 20 MeV/$c^2$ of the $D^0$ mass).

\subsection{A model for the $D^0 \to \pi^-\pi^+\pi^-\pi^+$ decay}

The decay mode $D^0\to\pi^-\pi^+\pi^-\pi^+$ is Cabibbo-suppressed and may proceed  through
many intermediate resonant states. 
Considering only tree-level amplitudes, possible contributions come from
two-body decays such as $D\to a_1(1260)^+\pi^-$, $D\to \rho(770)^0\rho(770)^0$, $D\to \rho(770)^0S$ and  $D\to SS$,
where $S$ is a scalar meson ($S=f_0(600)$ or $\sigma, ~f_0(980),~f_0(1370)$). The $\pi^-\pi^+\pi^-\pi^+$ can also
result from three-body nonresonant decays 
$D \to R\pi\pi$ ($R=f_0(600)$ or $\sigma, ~\rho(770)^0, ~f_0(980), ~f_2(1270), ~f_0(1370), ~\rho(1450)^0$).  
Including all possible contributions, one could have over twenty different resonant amplitudes leading to the  
$\pi^-\pi^+\pi^-\pi^+$ final state. 

In addition to the resonant modes, the  $\pi^-\pi^+\pi^-\pi^+$ final state could also result from a 
four-body nonresonant decay.
Usually the nonresonant component is assumed to be uniform, but this may not be a reasonable assumption
even in the simpler case of three-body decays \cite{bedcg}. To our knowledge there is no phenomenological model for
a non-uniform nonresonant amplitude in four-body decays.

The existence of many possible intermediate states leading to the $\pi^-\pi^+\pi^-\pi^+$ final state makes
the amplitude analysis of this decay very challenging. Many of these intermediate states involve broad resonances 
that populate the whole phase space. A model having a large number of overlapping amplitudes  gives rise to 
large interference terms which are difficult to control in a five-dimensional space. The presence of these 
interference terms is easily detected when the sum of decay fractions greatly exceeds 100\%. In such cases
many local minima exist in the parameter space, with very similar likelihood values. With such a model one may find a
mathematical solution to the fit problem, losing, however, the physical meaning of it.
An additional difficulty comes from the fact that in the $\pi^-\pi^+\pi^-\pi^+$ final state there are two pairs of
identical pions and therefore amplitudes must be Bose-symmetrized. In the chain $D^0 \to a_1(1260)^+ \pi^-$,
$a_1(1260)^+ \to \rho^0 \pi^+$, $\rho^0 \to \pi^+\pi^-$, for instance, one can combine the four
pions in eight possible ways. Angular distributions, a clean signature of specific modes, tend to be smeared
out by the Bose-symmetrization. This could be minimized using flavor tagging, since a large fraction of the
$D^0$ comes from the decay  $D^{*\pm} \to D^0 (\overline{D}{}^0)\pi^{\pm}$. In our case that would
reduce the  sample size by a factor of five. Finally, as the number of final state particles increases
the more important the final state interactions tend to be. The FSI may play a significant role in the
$D^0 \to \pi^-\pi^+\pi^-\pi^+$ decay. Unfortunately this is a very difficult problem to deal with, 
even with a phenomenological approach.

The strategy adopted in this analysis is to start with a set of amplitudes
corresponding to modes that are expected to be dominant. Once a stable solution is achieved, modes
with marginal contributions are replaced by other ones, until a final set is reached.

\begin{figure}
\includegraphics[width=8.5 cm]{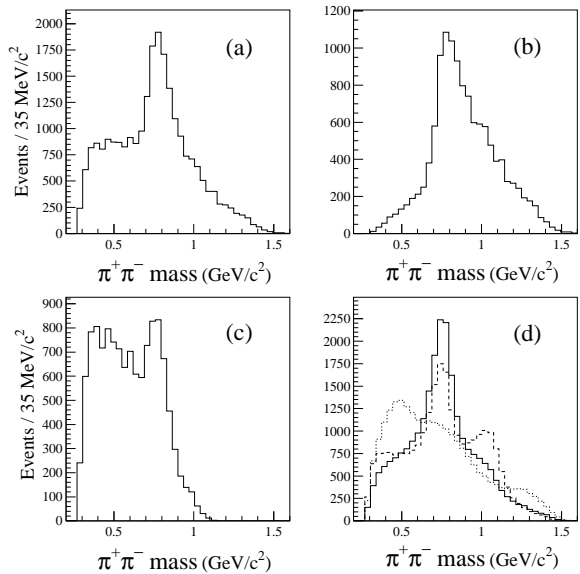}
\caption{Invariant mass distribution for $\pi^-\pi^+$.
(a) All four $\pi^-\pi^+$ combinations; (b) the $\pi^-\pi^+$ combinations with highest mass; 
(c) the $\pi^-\pi^+$ combinations with lowest mass;
(d) MC simulation of the $\pi^-\pi^+$ mass distribution for $D^0\to a_1(1260)^+\pi^-$: the solid line corresponds to
the $a_1^+ \to \rho(770)^0\pi^+$ S-wave, the dashed line is the $a_1^+ \to \rho(770)^0\pi^+$ D-wave, and the dotted 
line is $a_1^+ \to \sigma\pi^+$.}
\label{proj1}
\end{figure}

We built a baseline mode guided by the $\pi^-\pi^+$ projections, by MC simulation of the different channels and by our own
experience. The $\pi^-\pi^+$ mass projections are shown in Fig. \ref{proj1}. 
The plot in Fig. \ref{proj1}(a) has four entries per event, corresponding to the
four different $\pi^-\pi^+$ combinations one could form in each event.
We adopt the particle labeling $D^0 \to \pi^-_1\pi^+_2\pi^-_3\pi^+_4$, so the identical particles 
are $\pi_1/\pi_3$ and $\pi_2/\pi_4$. In our sample we do not distinguish between $D^0$ and $\overline{D}{}^0$,
and the like charge pions are randomized to avoid any ordering with respect to their momenta. Consequently,
all four pions have the same momentum distribution, making the four $\pi^-\pi^+$ mass projections 
indistinguishable.  

One can form two combinations of $\pi^-\pi^+$ pairs, $\pi^-_1\pi^+_2$/$\pi^-_3\pi^+_4$ 
or $\pi^-_1\pi^+_4$/$\pi^-_2\pi^+_3$. One can plot the highest and lowest $\pi^-\pi^+$ invariant mass
for each pair combination. These are shown in Fig. \ref{proj1}(b)-(c), respectively. In these plots there are 
two entries per event. 

The $\rho(770)^0$ is clearly the dominant $\pi^-\pi^+$ resonance, although other resonances,
like the $\sigma$, $f_0(980)$ and  $f_2(1270)$ may also be present, even without leaving a clear signature in the 
$\pi^-\pi^+$ mass projection. The $\rho(770)^0$ signal 
could originate from the decay of the axial-vector meson $a_1(1260)$, from the decay of the type
$D^0 \to \rho^0\rho^0$, $D^0 \to \rho^0 R$, or even from $D^0 \to \rho^0 \pi^+\pi^-$. 
Figure 2(d) shows the $\pi^-\pi^+$ projection of
Monte Carlo (MC) simulations of the decay $D^0 \to a_1(1260)^+\pi^-$, with the 
$a_1$ decaying to $\rho^0 \pi^+$ (S- and D-wave) and to $\sigma\pi^+$.

Our baseline model, therefore, includes contributions of three types. The first type is the 
$D^0 \to a_1(1260)^+\pi^-$ chain. For the $a_1(1260)$ resonant substructure, we consider three channels:  
$a_1^+ \to \rho(770)^0\pi^+$, the  $\rho^0\pi^+$ being in a dominant S-wave state with a small D-wave component, 
plus $a_1^+ \to \sigma \pi^+$. 
The second type of contribution is the $D^0 \to \rho^0\rho^0$ decay, in three possible helicity states. Finally,
we consider decays of the type $D^0 \to R\pi^+\pi^-$, with $R= \sigma, \rho^0, f_0(980)$ and 
$f_2(1270)$.

\subsection{Formalism}

The formalism used in this amplitude analysis is a straightforward extension to four-body decays 
of the usual Dalitz plot fit technique. The $D^0 \to \pi^-\pi^+\pi^-\pi^+$ signal amplitude 
is represented by a coherent sum of individual amplitudes, each corresponding to a possible intermediate 
state and weighted by constant complex coefficients $c_k$,  ${\mathcal A} = \sum c_k A_k$.  
The background amplitude is represented by an incoherent sum of amplitudes corresponding to the 
different types of background events.

A likelihood function is built from the signal and background amplitudes, incorporating
detector resolution effects and acceptance correction.
The optimum set of constants $c_k$, the only unknowns, is obtained by an unbinned maximum likelihood fit.
The background amplitude is kept fixed during the fit procedure. 

\subsubsection{Phase space}

The kinematics of the $D^0 \to \pi^-\pi^+\pi^-\pi^+$ decay is unambiguously determined by five degrees of freedom. 
We choose a set of five two-body invariant masses squared
--- the four different $\pi^-\pi^+$ combinations, $s_{12},s_{14},s_{23},s_{34}$ plus $s_{13}$ --- to define an 
event in the five-dimensional phase space. Any kinematic variable, like acoplanarity and helicity angles 
or three-body masses, can be easily expressed in terms of this set of invariants.
Unlike the case of three-body decays, the phase space density is no longer constant.

\subsubsection{Signal amplitudes}

The amplitudes for the intermediate states, $A_i$, are phenomenological objects. Each signal amplitude is 
built as a product of relativistic Breit-Wigner functions (BW), an overall spin amplitude, $\mathcal{M}$, accounting 
for angular momentum conservation at each decay vertex, and form factors, $F_l$ ($l$ is the orbital angular 
momentum of the decay vertex), accounting for the finite hadron size. This is the standard structure for 
resonant amplitudes (see note of Dalitz plot formalism in \cite{pdg}). The amplitude for the decay chain 
$D^0 \to a_1(1260)^+ \pi^-$, $a_1(1260)^+ \to \rho^0 \pi^+$ (S-wave), $\rho^0 \to \pi^+\pi^-$, is 

\begin{equation}
A=F^D_1 \times \mathrm{BW}^{a_1} \times F^{a_1}_1 \times \mathrm{BW}^{\rho} \times F^{\rho}_1 \times \mathcal{M}.
\end{equation}

The amplitude for the decay chain $D^0 \to \rho^0_1 \rho^0_2$, $\rho^0_1  \to \pi^+ \pi^-$, 
$\rho^0_2 \to \pi^+ \pi^-$ is

\begin{equation}
A=F^D_l \times \mathrm{BW}^{\rho_1} \times F^{\rho_1}_1 \times \mathrm{BW}^{\rho_2} \times F^{\rho_2}_1   
\times \mathcal{M},
\end{equation}
and for the chain $D^0 \to f_2(1270) \pi^+ \pi^-$, $f_2(1270) \to \pi^+ \pi^-$ the amplitude is

\begin{equation}
A=F^D_2 \times \mathrm{BW}^{f_2} \times F^{f_2}_2 \times \mathcal{M}.
\end{equation}

All signal amplitudes are Bose-symmetrized due to the existence of two pairs of identical particles in the final
state.

We use the Blatt-Weisskopf damping factors \cite{blatt}  as form factors for vertices involving spin-1 and spin-2 
resonances. These form 
factors have one free parameter, $r_R$, which is related to the resonance size. 
All form factors are slowly varying functions of energy. As a systematic check we varied the value of $r_R$.
The fit result is not very sensitive to the value of this parameter in the range $1<r_R<5 ~\mathrm{GeV}^{-1}$.
The form factor parameter is fixed at $r_R=3.0$  $\mathrm{GeV}^{-1}$.  

For the spin amplitudes we use the Lorentz invariant amplitudes for the
decay sequences $D^0 \to a_1^+\pi^-$ and $D \to R\pi^+\pi^-$ ($R$ being a vector or tensor resonance).
These Lorentz invariant amplitudes are built using the 
relativistic tensor formalism \cite{filippini,chung} which combines particle 4-momenta with polarization 
vectors in a rotationally invariant way. In vertices involving only the strong interactions, parity conservation 
imposes an additional constraint. For the decay $D^0 \to \rho^0\rho^0$ the spin amplitudes
are written using the helicity formalism. Explicit formulae for the angular distributions  {\cal M} are 
presented in the Appendix.

The relativistic Breit-Wigner representing the $f_2(1270)$ and $\sigma$ resonances has an energy dependent width,
with the approximation $\Gamma_{\pi\pi}(s)=\Gamma_{\mathrm{\mathrm{tot}}}(s)$,

\begin{equation}
\mathrm{BW} =  \frac{1}{s - s_0 + i\sqrt{s_0}\Gamma_{\mathrm{\mathrm{tot}}}(s)},
\label{bw}
\end{equation}
where

\begin{equation}
\Gamma_{\mathrm{\mathrm{tot}}}(s) = \Gamma_0 
\sqrt{\frac{s_0}{s}}\left(\frac{p^*}{p^*_0}\right)^{2l+1}\frac{F_l^2(p^*)}{F_l^2(p^*_0)}.
\end{equation}

In the above equations $s$ is the $\pi^+\pi^-$ mass squared, $p^*$ is the break-up momentum of the resonance,
$s_0$ the resonance nominal mass squared, and $p^*_0=p^*(s_0)$.

The $\rho^0$ line shape is better described when the interference with the $\omega(782)$ is included.
In spite of the tiny branching fraction of the $\omega(782) \to \pi^+\pi^-$, the effect of the
$\rho$--$\omega$ interference is remarkable and clearly distorts the pure $\rho^0$ line shape, as one can see by 
comparing the $\pi^+\pi^-$ with the $\pi^0\pi^{\pm}$ mass distributions from \cite{cbarrel2}. In this analysis
we use the line shape given by the Crystal Barrel Collaboration \cite{cbarrel1}.

For the $f_0(980)$ we used the Flatt\'e formula~\cite{flatte} of a coupled channel Breit-Wigner 
function. The $f_0(980)$ parameters --- $g_{\pi} = 0.20 \pm 0.04$, and
$g_K = 0.50 \pm 0.20$ and $m_0 = (0.957 \pm 0.008) ~\mathrm{GeV}/c^2$ --- are obtained 
by a fit to the FOCUS $D_s^+ \to \pi^+ \pi^- \pi^+$ Dalitz plot, 
where the $f_0(980) \pi^+$ is the dominant component.

The $f_2(1270)$ Breit-Wigner parameters are fixed at the PDG values \cite{pdg}, whereas for the $\sigma$
we used the E791 parameters \cite{e791}. It is well known that the $\sigma$ pole cannot be
obtained from a simple Breit-Wigner \cite{meissner}. Nevertheless, good fits have been obtained by
different experiments, in different channels, using  Breit-Wigners with similar parameters.
There is no consensus on the correct way to parameterize this low mass, broad scalar state.
The focus here, however, is not on the $\sigma$ properties, but
on the $a_1(1260)$ line shape, which depends on its resonant substructure. For this purpose it suffices to
use the Breit-Wigner formula as an effective representation of a scalar component of the $a_1(1260)$
resonant substructure. In Section VI.D we  discuss the sensitivity of the fit 
results to the values of the $\sigma$ parameters.

\subsubsection{The $a_1(1260)$ line shape}

One of the main goals of this analysis is to determine the $a_1(1260)$ line shape and  
its resonant substructure. These are correlated: we need to know the resonant substructure in order to
determine the line shape because the resonant substructure defines how the $a_1(1260)$ width,  
$\Gamma_{\mathrm{tot}}^{a_1}(s)$ depends on the $\pi^+ \pi^- \pi^+$ mass squared $s$. Given the
functional form of $\Gamma_{\mathrm{tot}}^{a_1}(s)$, we represent the $a_1$ line shape by the same
Breit-Wigner formula of Eq. (\ref{bw}). The total  width of the $a_1(1260)$ is given by

\begin{equation}
\Gamma_{\mathrm{tot}}^{a_1}(s) =  \Gamma_{2\pi^0\pi^+}^{a_1}(s) + \Gamma_{2\pi^+\pi^-}^{a_1}(s) + 
g^2_{K^*K}\Gamma_{K^*K}^{a_1}(s).
\end{equation}

We assume that $\Gamma_{2\pi^0\pi^+}^{a_1}(s) = \Gamma_{2\pi^+\pi^-}^{a_1}(s)$. The $a_1$ can also decay to
$KK\pi$ via $f_0(980)\pi$ or $K^*(892)K$. The opening of these channels introduces a cusp-like effect in the
$a_1$ width. Since no evidence of the mode $f_0(980)\pi$ has been reported so far, we
assume the $KK\pi$ partial width is entirely due to $K^*(892)K$. The value of the coupling 
constant $g^2_{K^*K}$ is taken from the analysis of the decay $\tau \to \pi^0\pi^0\pi^+\nu_{\tau}$ by
CLEO \cite{cleotau}. The coupling constant $g^2_{K^*K}$ was also measured by the Belle Collaboration
\cite{belle2}, which found a value three to five times larger than that measured by CLEO. The uncertainty
on the value of $g^2_{K^*K}$ is a source of systematic uncertainty.

The partial width
$\Gamma_{2\pi^+\pi^-}^{a_1}(s)$ is obtained by integrating the $\pi^+\pi^-\pi^+$ Dalitz plot,

\begin{equation}
\Gamma_{2\pi^+\pi^-}^{a_1}(s) \propto \frac{1}{s^{3/2}} \int ds_1ds_2 \left|{\cal A}_{a_1}(s_1,s_2) \right|^2,
\label{gamma} 
\end{equation}
with $s_1,s_2$ being the invariant mass squared of the two possible $\pi^+\pi^-$ combinations and

\begin{equation}
{\cal A}_{a_1}(s_1,s_2)=\sum_{i=1}^3 c_i A_i(s_1,s_2),
\end{equation}
where $c_i$ are complex coefficients and $A_i(s_1,s_2)$ the amplitudes for each of the $a_1$ decay modes.
We fix the partial width by demanding that $\Gamma_{2\pi^+\pi^-}^{a_1}(s=s_0^{a_1}) = \Gamma_0^{a_1}$,
with $s_0^{a_1}$ and $\Gamma_0^{a_1}$ being real input parameters to be determined.

In practice, we used an iterative procedure: assuming initial values for $s_0^{a_1}$ and $\Gamma_0^{a_1}$, 
and for the constants $c_i$, the total width $\Gamma_{\mathrm{tot}}^{a_1}(s)$ is computed. 
Having computed $\Gamma_{\mathrm{tot}}^{a_1}(s)$, we perform the five-dimensional fit, which returns 
new values for the constants $c_i$; the function $\Gamma_{2\pi^+\pi^-}^{a_1}(s)$ is then recomputed using the 
same values of $s_0^{a_1}$ and $\Gamma_0^{a_1}$ and the new values of the constants $c_i$. The process converges after
a few iterations. The whole procedure is repeated many times, scanning over the values of $s_{a_1}$ and $\Gamma_0^{a_1}$, 
until we find the combination that optimizes the fit result.

\subsubsection{Background}

The background composition was determined by inspection of the data on the side band to the right 
of the $D^0$ signal in the $\pi^-\pi^+\pi^-\pi^+$ mass spectrum.
We consider two types of background events: random combinations of a $\rho(770)^0$ and a
$\pi^-\pi^+$ pair and random combinations of $\pi^-\pi^+\pi^-\pi^+$. 
We estimate a fraction of 80\% for random combinations of $\pi^-\pi^+\pi^-\pi^+$ and 20\% for 
$\rho(770)^0$ plus a $\pi^-\pi^+$ pair. Random combinations of two $\rho^0$ might be also present, although with a
small contribution. The background fractions, as well as its composition, is a source of systematic uncertainty.

We assume the random  
$\pi^-\pi^+\pi^-\pi^+$ combinations to be uniformly distributed in phase space, whereas
for the other background we assume a Breit-Wigner with no
form factors and no angular distribution.
The overall background distribution is a weighted, incoherent 
sum of the two components described above. The relative background fractions, 
$b_k$, are fixed in the fit.

\subsubsection{Detector resolution and acceptance correction}

The finite momentum resolution causes a smearing of the phase space boundary. It is easier to
understand this effect if we consider the case of a three-body decay and the Dalitz plot. 
In a sample where the three-body masses have only one well defined value, the Dalitz 
plot boundary is also  well defined and unique. But in a sample where we have a distribution of
three-body masses, $G(M)$, the Dalitz plot boundary is no longer well defined but, rather, it 
is a superposition of boundaries, each one defined by its corresponding value of the three-body mass. 
Dividing the three-body mass distribution into fine bins, one can think of the observed 
$s_{12} ~\times ~s_{23}$ distribution  as a superposition of 
Dalitz plots, each one weighted by $G(M_i)$. The same reasoning applies to four-body decays. We
incorporate the effect of the momentum resolution, both in the signal and background probability distribution
functions (PDF), by weighting each event by the 4$\pi$ mass distribution. With this 
approach, the signal and background PDF depend on the five invariants and on the 4$\pi$ mass as well. 

The acceptance correction is also applied on an event-by-event basis. The acceptance is a function
of the five invariants. In the case of the amplitude analysis,
the absolute value of the acceptance function is irrelevant, since it implies an overall constant factor 
multiplying  the likelihood. All one really needs to know is the acceptance at a given phase space cell 
relative to its neighbor cells. We use 
a five-dimensional matrix for the acceptance correction rather than a continuous function. We used the
full FOCUS Monte Carlo simulation (ROGUE) to generate a very large sample of $D^0 \to \pi^-\pi^+\pi^-\pi^+$
events with a constant matrix element. We applied the same selection criteria to this sample as those used 
for real data.
A mini-MC simulation of the $D^0 \to \pi^-\pi^+\pi^-\pi^+$ was used to generate the phase space distribution.
Each of the five axes were divided into bins. Two five-dimensional arrays were filled, one with the events of
the full simulation that passed all cuts and the other with the phase space events. The acceptance matrix
was formed by dividing the number of ROGUE events in each cell by the number of phase space events in that cell.
The acceptance is nearly uniform across the phase space, with a 10-15\% decrease close to the edges.

\subsubsection{Normalization}

The maximum likelihood fit technique requires the likelihood function to be normalized. This means one
normalization constant for the signal and another for the background probability distribution functions.
The normalization constant for the background distribution needs to be calculated just once,
since the background amplitudes and the relative fractions are fixed during the fit. The
normalization constant for the signal, however, must be computed at each step of the minimization
procedure, since it involves the fit parameters.
 
Although it is not necessary, the amplitudes for each intermediate state are normalized to unity. 
This is intended to give the magnitudes $c_k$ a direct physical meaning: the decay fraction of each mode 
is directly proportional to $|c_k|^2$. 

The normalization is a crucial step in amplitude analysis. The evaluation of the normalization constants
requires a number of phase space integrals. Fortunately this integration can be performed just once, 
since  all input parameters in the amplitudes --- mass and width of resonances, form factors parameter --- are fixed 
during the minimization process.

We define
$G(m_{4\pi})$ and $b(m_{4\pi})$ as functions representing the 4$\pi$ mass signal and background 
distributions, respectively. The function $G(m_{4\pi})$ is a sum of two Gaussians, whereas 
$b(m_{4\pi})$ is a sum of two exponentials. Both $G(m_{4\pi})$ and $b(m_{4\pi})$ were 
obtained by the fit to the $4\pi$ mass spectrum shown in Fig.~\ref{m4pi}. If we define $\phi$ as
a point in phase space ($d\phi=ds_{12}ds_{14}ds_{23}ds_{34}ds_{13}$), $\rho(\phi)$ as the phase space 
density, and $\varepsilon(\phi)$ as the efficiency, and $\bar A_k(\phi)$, $\bar B_j(\phi)$ the unormalized amplitudes,
the normalization integrals are

\begin{equation}
N_k^S = \int dm_{4\pi} ~G(m_{4\pi}) \int d\phi ~\rho(\phi) \left|\bar A_k(\phi) \right|^2
\end{equation}
and

\begin{equation}
N_j^B = \int dm_{4\pi} ~b(m_{4\pi}) \int d\phi ~\rho(\phi) \bar B_j(\phi).
\end{equation}

The overall signal and background normalization integrals are

\begin{equation}
N_S = \int dm_{4\pi} ~G(m_{4\pi}) \int d\phi ~\rho(\phi) \varepsilon(\phi) 
\left|\sum c_k A_k(\phi) \right|^2
\end{equation}
and

\begin{equation}
N_B = \int dm_{4\pi} ~b(m_{4\pi}) \int d\phi ~\rho(\phi) \varepsilon(\phi) \sum b_j B_j(\phi),
\end{equation}
with $A_k(\phi)=\bar A_k(\phi)/N_k^S$ and $B_j(\phi)=\bar B_j(\phi)/N_j^B$.

We performed the phase space integration using the Monte Carlo. The $m_{4\pi}$ interval is divided into bins.
For each $m_{4\pi}$ bin a very large sample of MC events, generated with a constant matrix
element, is used to compute average value of the integrands. These average values are then multiplied by the phase space
volume and weighted by $G(m_{4\pi})$/$b(m_{4\pi})$.

\subsubsection{The likelihood function}

The overall signal and background amplitudes are corrected on an event-by-event basis 
for the acceptance, which is nearly constant across the phase space, and for the finite detector 
resolution, taken into account by multiplying the overall signal distribution 
by a Gaussian factor, $G(m_{4\pi})$, and the background distribution by an exponential function, 
$b(m_{4\pi})$. The normalized signal probability distribution is, thus,

\begin{equation}
S_{\mathrm{PDF}} (\phi;m_{4\pi}) = \frac{1}{N_S}  G(m_{4\pi}) \left|\sum c_k A_k(\phi) \right|^2,
\label{spdf}
\end{equation}
and
 
\begin{equation}
P_S(\phi;m_{4\pi}) = \varepsilon(\phi) \rho(\phi) S_{\mathrm{PDF}} (\phi;m_{4\pi}).
\end{equation}

The normalized background probability distribution is

\begin{equation}
B_{\mathrm{PDF}} (\phi;m_{4\pi}) = \frac{1}{N_B} b(m_{4\pi}) \sum b_k B_k(\phi),
\label{bpdf}
\end{equation}
and

\begin{equation}
P_B(\phi;m_{4\pi}) = \varepsilon(\phi) \rho(\phi) B_{\mathrm{PDF}} (\phi;m_{4\pi}). 
\end{equation}

An unbinned maximum likelihood fit was performed, minimizing the quantity
$w\equiv -2\log(\mathcal{L})$. The likelihood function, $\mathcal{L}$, is

\begin{equation}
\mathcal{L} = \prod_\mathrm{events} \left[P_S(\phi^i;m_{4\pi}^i) + P_B(\phi^i;m_{4\pi}^i)\right] = 
\prod_\mathrm{events} \left[\varepsilon(\phi^i) \rho(\phi^i) (S_{\mathrm{PDF}} (\phi^i;m_{4\pi}^i) + 
B_{\mathrm{PDF}} (\phi^i;m_{4\pi}^i))\right]
\end{equation}

When we take the logarithm of the likelihood we have

\begin{equation}
\log \mathcal{L} = \sum_\mathrm{events} \log\left[\varepsilon(\phi^i) \rho(\phi^i)\right] + 
\sum_\mathrm{events} \log\left[S_{\mathrm{PDF}} (\phi^i;m_{4\pi}^i) + B_{\mathrm{PDF}} (\phi^i;m_{4\pi}^i)\right]
\end{equation}

In the fit parameter space the term $\sum \log\left[\varepsilon(\phi^i) \rho(\phi^i)\right]$ is a constant and does
not affect the position of the minimum. In practice, thus, we minimize the quantity 

\begin{equation}
w=-2 \left[ \sum_\mathrm{events} \log \left(\frac{1}{N_S}  G(m_{4\pi}) \left|\sum c_k A_k(\phi^i) \right|^2
+ \frac{1}{N_B} b(m_{4\pi}) \sum b_k B_k(\phi^i)\right)\right] .
\end{equation}

The acceptance correction and the phase space density 
enter only in the overall normalization constants $N_S$ and $N_B$.

Decay fractions are obtained from the coefficients $c_k$, determined by the fit,
and after integrating the overall signal amplitude over the phase space at $m_{4\pi}=m_{D^0}$,

\begin{equation}
f_k = \frac{\int d\phi\left|c_k A_k(\phi) \right|^2}{\int d\phi \left|\sum_j c_j A_j(\phi) \right|^2} = 
\frac{\left|c_k \right|^2}{\int d\phi \left|\sum_j c_j A_j(\phi) \right|^2},
\end{equation}
since the individual amplitudes $A_k(\phi)$ are normalized to unity.
Errors on the fractions include errors on both magnitudes and phases, and are computed using the 
full covariance matrix.

\section{Results from amplitude analysis}

The technique described in Section V is applied to the events in the hatched area of Fig. 1. 
The decay $D^0 \to a_1^+\pi^-$, $a_1^+ \to \rho^0 \pi^+$ (S-wave) is taken as the reference mode, fixing the phase
convention and the relative magnitudes of the other contributions. The results of the best fit to the data
are summarized in Table \ref{fractions}. The systematic errors are discussed in Section VI.E.

\begin{table}
\caption{Results from the best fit. The first error is statistical, and the second one is systematic.}
\label{fractions}
\begin{ruledtabular}
\begin{tabular}{lccr}
  	                   mode       &      magnitude                & phase (degrees)     &	fraction (\%)
\\ \hline 
$a_1^+\pi^-, ~a_1 \to \rho^0 \pi^+$  ~(S-wave)& 1. (fixed)                    &  0 (fixed)          & 43.3 $\pm$ 2.5 $\pm$ 1.9
 \\ 
$a_1^+\pi^-, ~a_1 \to \rho^0 \pi^+$  ~(D-wave)& 0.241 $\pm$ 0.033 $\pm$ 0.024 &  82 $\pm$ 5 $\pm$ 4 &  2.5 $\pm$ 0.5 $\pm$ 0.4
 \\ 
$a_1^+\pi^-, ~a_1 \to \sigma \pi^+$         & 0.439 $\pm$ 0.026 $\pm$ 0.021 & 193 $\pm$ 4 $\pm$ 4 &  8.3 $\pm$ 0.7 $\pm$ 0.6
 \\
$a_1^+\pi^-  ~(\mathrm{all})$             & -                             &  - 	            & 60.0 $\pm$ 3.0 $\pm$ 2.4
 \\ 
$\rho^0\rho^0   ~(\mathrm{parallel}) $    & 0.157 $\pm$ 0.027 $\pm$ 0.020 & 120 $\pm$ 7 $\pm$ 8 &  1.1 $\pm$ 0.3 $\pm$ 0.3
 \\ 
$\rho^0\rho^0   ~(\mathrm{perpendicular})$& 0.384 $\pm$ 0.020 $\pm$ 0.015 & 163 $\pm$ 3 $\pm$ 3 &  6.4 $\pm$ 0.6 $\pm$ 0.5
 \\ 
$\rho^0\rho^0   ~(\mathrm{longitudinal}) $& 0.624 $\pm$ 0.023 $\pm$ 0.015 & 357 $\pm$ 3 $\pm$ 3 & 16.8 $\pm$ 1.0 $\pm$ 0.8
 \\ 
$\rho^0\rho^0   ~(\mathrm{all})$          & -                             & -  	            & 24.5 $\pm$ 1.3 $\pm$ 1.0
 \\ 
$f_0(980) \pi^+ \pi^- $                   & 0.233 $\pm$ 0.019 $\pm$ 0.015 & 261 $\pm$ 7 $\pm$ 4 &  2.4 $\pm$ 0.5 $\pm$ 0.4
 \\  
$f_2(1270) \pi^+\pi^-$                    & 0.338 $\pm$ 0.021 $\pm$ 0.016 & 317 $\pm$ 4 $\pm$ 4 &  4.9 $\pm$ 0.6 $\pm$ 0.5
 \\ 
$\sigma \pi^+\pi^-$                       & 0.432 $\pm$ 0.027 $\pm$ 0.022 & 254 $\pm$ 4 $\pm$ 5 &  8.2 $\pm$ 0.9 $\pm$ 0.7
 \\
$R\pi^+\pi^-$   ~(all)                    & -                             & -  	            & 20.0 $\pm$ 1.2 $\pm$ 1.0
\end{tabular}
\end{ruledtabular}
\end{table}

\begin{figure}
\includegraphics[width=10.5 cm]{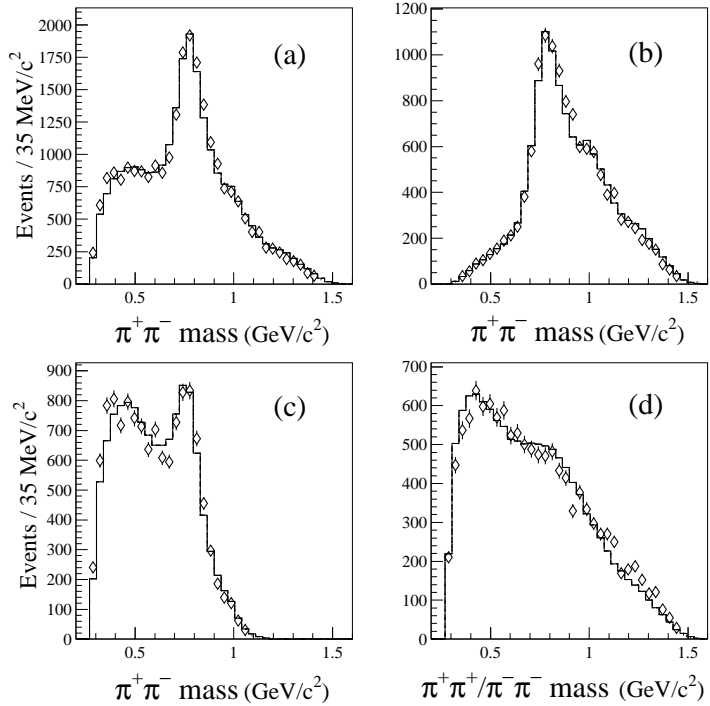}
\caption{Comparison between data and the result of the best fit. 
In (a) all four $\pi^-\pi^+$ combinations are added. Entries from plot (a) are split into
combinations with highest and lowest  $\pi^-\pi^+$ mass: in (b) we show  the $\pi^-\pi^+$ combinations 
with highest mass, whereas in (c) we show the $\pi^-\pi^+$ combinations with lowest mass; finally, in
(d) we show the $\pi^-\pi^-/\pi^+\pi^+$ mass distribution. In all plots the solid histogram is a
projection of the fit.}
\label{proj2}
\end{figure}

In Fig. \ref{proj2} the $\pi\pi$ mass projections from data (diamonds with error bars) is plotted
with the fit result overlaid (solid histograms). The  fit projections  are made from a large 
MC simulated sample, including the signal and background PDF's, defined in Eq. (\ref{spdf}) and 
(\ref{bpdf}), as well as the reconstruction efficiency and detector resolution. In Fig. \ref{proj2}(a) the
$\pi^-\pi^+$ mass is shown. Events in Fig. \ref{proj2}(b) and (c) are the $\pi^-\pi^+$
combinations with highest and lowest mass, respectively. In Fig. \ref{proj2}(d) we show the 
$\pi^-\pi^-$/$\pi^+\pi^+$ invariant mass.

\begin{figure}
\includegraphics[width=8.5 cm]{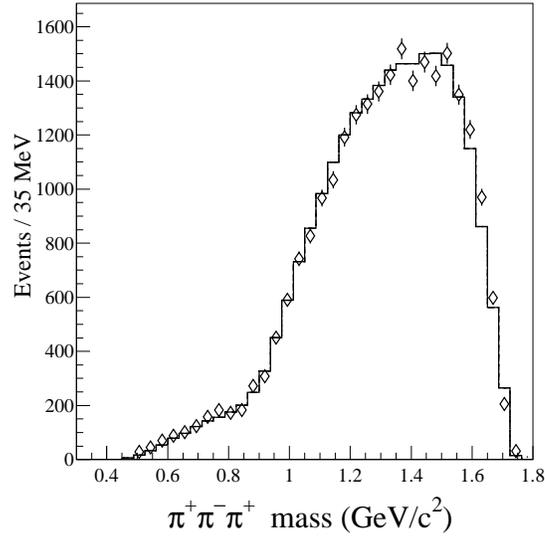}
\caption{The $\pi^+\pi^-\pi^+$/$\pi^-\pi^+\pi^-$ mass spectrum. A projection of the fit (solid histogram) is
overlaid with the data points (diamonds with error bars).}
\label{m3pi}
\end{figure}

The $\pi^+\pi^-\pi^+$ mass spectrum is shown in Fig. \ref{m3pi}. This plot has four entries per event.
The diamonds with error bars are the data distribution, whereas the solid histogram is the fit projection. 
Note that this distribution does not reflect the pure $a_1(1260)$ line shape, for, in addition to the
Bose-symmetrization, there are contributions from the other modes.

\begin{figure}
\includegraphics[width=8.5 cm]{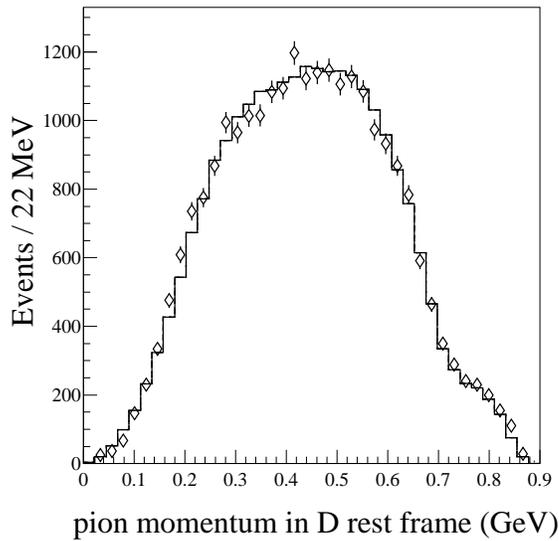}
\caption{The pion momentum distribution in the $D$ rest frame (four entries per event). The momentum 
distribution for
all pions should be the same, so in this plot there are four entries per event. A projection of the fit 
(solid histogram) is
overlaid with the data points (diamonds with error bars).}
\label{mom}
\end{figure}

A further comparison between the fit result and the data can be made with the pion momentum distribution, 
computed in the $D$ rest frame. Recall that we do not distinguish between $D^0$ and $\overline{D}{}^0 $, and that
there is no particle ordering according to its momentum. This means that all pions must have the same 
momentum distribution and that we can add them into a single plot. The pion momentum distribution is
shown in Fig. (\ref{mom}), where the diamonds represent the data points, and the solid histogram is a 
MC simulation.

\begin{figure}
\includegraphics[width=8.5 cm]{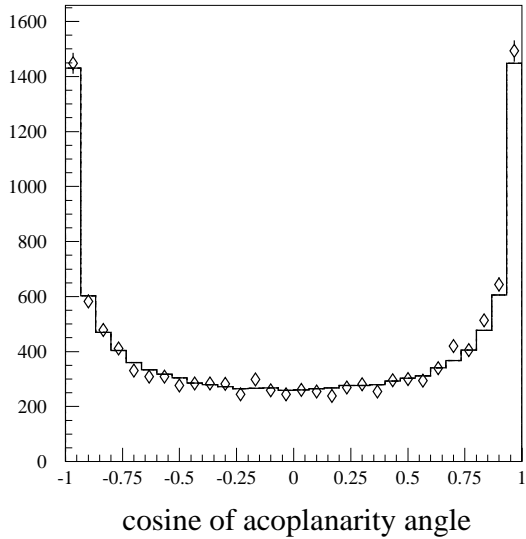}
\caption{The cosine of the acoplanarity angle distribution. The contribution of 
the $D^0 \to \rho^0\rho^0$ is only 25\%, so the original polarization of the two vector states is diluted.}
\label{ang}
\end{figure}

We can also compute the acoplanarity angle. In the $D \to RR$ decay this is the angle, measured in the $D$ 
rest frame, between the planes defined by the decay particles of each of the resonances. Note that in the 
$D^0 \to \pi^-\pi^+\pi^-\pi^+$ decay the $\rho^0\rho^0$ mode accounts for only one fourth of the decay rate, 
so the distribution of the acoplanarity angle no longer carries the information on the polarization state 
of the two vector mesons. The distribution cosine of the acoplanarity angle is shown in Fig. \ref{ang}.
Once more, the diamonds represent the data points, and the solid histogram is a projection of the fit.

\subsection{Goodness-of-fit}

The $\pi^+\pi^-$ and the $\pi^+\pi^-\pi^+$ mass spectra, as well as the pion momentum and the cosine of the
acoplanarity angle distribution, are reasonably well represented by our model. These are, however, inclusive distributions,
indicating that the model represents the general features of the data. Any quantitative statements about 
the fit quality should be made by inspecting the five-dimensional phase space. 

The goodness-of-fit is assessed using a $\chi^2$-like test. The phase space is divided in equally-sized,
five-dimensional cells. For each of these, the expected number of events, $n_{\mathrm{fit}}$, is obtained by the 
MC simulation described above, after scaling the MC sample to the number of events in the data sample.
For each cell with a minimum expected occupancy of 5 events, a $\chi^2$ is computed comparing  the 
predicted population to the observed number of events, $(n_{\mathrm{obs}}-n_{\mathrm{fit}})/\sigma$, where 
$\sigma=\sqrt{n_{\mathrm{obs}}}$. The total $\chi^2$ is formed by summing over all cells with minimum 
occupancy. The number of degrees of freedom is the number of cells with minimum occupancy minus the 
number of free parameters in the fit. The baseline model has 9 amplitudes, with 16 free parameters. 
There are 166 bins with at least the minimum occupancy, so the number of degrees of freedom is 150.
These bins include 97\% of the data. The total $\chi^2$ is 348.0, 
which corresponds to a confidence level (CL) of $10^{-17}$. We conclude from the $\chi^2$ test that the baseline model 
does not provide an accurate description of the data in the five-dimensional phase space. 

We have looked at the distribution of the $\chi^2$ throughout the phase space. That might reveal the need of 
a missing resonant amplitude, or problems with the representation of the existing amplitudes. We found that 
there are 33 bins with $\chi^2>6$, containing approximately 15\% of the data events. The bins that are responsible
for the poor confidence level of the fit are evenly distributed throughout the phase space, 
instead of being concentrated 
in specific regions. In general, these 33 bins are surrounded by bins having a good $\chi^2$. 

Addition of further amplitudes of the type
$D \to RR$ do not improve the fit quality. The inclusion of an uniform nonresonant term gives rise to large 
interference with the $\sigma\pi^+\pi^-$ mode without improving the likelihood. A fit using an alternative 
model, in which the $D \to \sigma\pi^+\pi^-$ mode is replaced by a uniform  nonresonant amplitude, is worse 
than our best fit by more than 200 units of $w=-2\log{\cal L}$. No improvement is obtained when other amplitudes of
the type $D \to R\pi^+\pi^-$ are added. The contribution from $D^0 \to \rho^0\pi^+\pi^-$ is negligible. This is also
the case for amplitudes with higher mass states, like the $\rho(1450)^0\pi^+\pi^-$ and the $f_0(1370)\pi^+\pi^-$.

It is very unlikely that the poor CL is caused by problems with the representation of the signal amplitudes. 
The dominant resonance is the $\rho(770)^0$, which has a very well determined line shape. 
As we will discuss on Section VI.E, the uncertainty in the
line shape of the S-wave components does not have a large impact on the fit result. The other ingredients of the
signal amplitudes are the standard angular distributions and the well known, widely used Blatt-Weisskopf form factors,
depending on one single parameter whose value does not significantly affect the fit result. 

The isobar model is a very standard tool, and has been successfully used for several decades. Almost all 
Dalitz plot analyses of heavy flavor decays were performed by representing the signal distribution as a
coherent sum of resonant amplitudes with constant coefficients. The isobar model has also  been successful in 
describing four-body decays in the past \cite{mkiii,e691} (although no CL was quoted in these analyses)
and recently \cite{kkkp,kkpp}, but in all cases the 
data samples have limited statistics. This analysis is the first attempt to use the isobar model for an
amplitude analysis of a four-body decay with moderately high statistics. The underlying picture of the isobar model 
may be too simplistic, signaling other types of effects should be taken into account.

\subsection{General features of the best solution}

The dominant contribution to the $D^0 \to \pi^- \pi^+ \pi^- \pi^+$ decay comes from the $D^0 \to a_1(1260)^+\pi^-$ mode, 
accounting for 60\% of the decay rate. The second most dominant contribution is the $D^0 \to\rho(770)^0\rho(770)^0$ mode,
with a decay fraction of 25\%. Modes of the type $D \to R\pi^+\pi^-$ correspond to 20\% of the decay rate.

The large value of the $D^0 \to a_1(1260)^+\pi^-$ decay fraction supports the picture of an external
$W$-radiation amplitude, with  the virtual $W$  coupling to the $a_1(1260)$, as the dominant
mechanism for the  $\pi^-\pi^+\pi^-\pi^+$ final state. The same dominance has also been  observed in other
four-body decays, such as $D^0 \to K^-\pi^+\pi^-\pi^+$ and  $D^+ \to \overline{K}{}^0\pi^+\pi^-\pi^+$ \cite{mkiii,e691}. 
The same picture can be drawn in the case of three-body decays with the $W$ coupling to the $\rho(770)^+$, such as
$D^0 \to  K^-\pi^+\pi^0$ \cite{e687b} or
$D^+ \to  \overline{K}{}^0 \pi^+\pi^0$ \cite{mkiiib}, where the contribution from $D \to K\rho$ exceeds 50\%. 
This pattern is similar to the well known vector-dominance in electromagnetic interactions. It
can be understood as a manifestation of the V-A nature of the weak interaction.

The relatively large contribution (25\%) of the  $D \to \rho^0\rho^0$ ($D \to V^0_1V^0_2$)  decay has also been seen 
in other final states, such as $D^0 \to K^-\pi^+\pi^-\pi^+$, where the mode $D^0 \to \overline{K}{}^{*0}\rho^0$ accounts for
17\% of the decay rate, according to \cite{mkiii} or over 40\%, according to \cite{e691}. In the case of
the final state $D^0 \to K^-K^+\pi^-\pi^+$ \cite{kkpp}, the contribution of the modes
$D^0 \to \rho^0\phi$ and $D^0 \to \overline{K}{}^{*0}K^{*0}$ amounts to more than 30\%. 
In all cases the dominant $D \to V^0_1V^0_2$ 
amplitude is the internal $W$-radiation, which is expected to be suppressed with respect to
the external $W$-radiation amplitude.

The remainder of the decay rate is due to three-body nonresonant modes $D \to R\pi^+\pi^-$. 
Like the $D \to \rho^0\rho^0$ decay, the  internal $W$-radiation should be the dominant amplitude.
For any possible meson $R$ there are several ways to form a $R\pi^+\pi^-$ state with $J=0$,
combining the orbital angular momentum between the two pions from the nonresonant $\pi^+\pi^-$ system, the  
orbital angular momentum between this system and the resonance $R$, and the spin of the resonance. Consequently,
for any possible resonance $R$ we can assign different spin amplitudes.
We tried all possible assignments for the $\rho(770)^0\pi^+\pi^-$ and for the $\rho(1450)^0\pi^+\pi^-$ spin amplitudes,
but none yielded a significant contribution.
It is interesting to note that the only $D \to R\pi^+\pi^-$ modes with a significant contribution are 
$\sigma \pi^+ \pi^- $,  $f_0(980) \pi^+ \pi^-$ and $f_2(1270) \pi^+ \pi^- $, with the nonresonant $\pi^+\pi^-$ pair
being in a pseudoscalar state. The $\pi^+\pi^-$ pair and the resonance must then be in a state of even relative orbital 
angular momentum, and the resulting spin amplitudes coincide with the ones from three-body $D$ decays \cite{pdg}.

\subsection{$a_1(1260)$ results}

As we mentioned before, the study of the $a_1(1260)$ is one of the main purposes of this analysis.
The $a_1$ line shape is connected to the resonant substructure through Eq. (\ref{gamma}). 
The data seem to require, in addition to the dominant $a_1^+ \to \rho(770)^0\pi^+$ amplitude, an S-wave
component $a_1^+ \to \sigma\pi^+$, as in the case of CLEO \cite{cleotau} and E852 \cite{e852}. 
A fit without the $\sigma\pi^+$ mode increases the quantity $w=-2\log{\cal L}$ by more than 200.
We tried other modes, such as the $a_1^+ \to f_0(980)\pi^+$, $a_1^+ \to \rho(1450)^0\pi^+$ and 
$a_1^+ \to f_0(1370)\pi^+$, but their
contribution is negligible. We have also made a fit with the $a_1^+ \to f_0(980)\pi^+$ and 
$a_1^+ \to f_2(1270)\pi^+$ amplitudes replacing the $f_0(980)\pi^+\pi^-$ and $f_2(1270)\pi^+\pi^-$. 
This fit is worse by more than 100 units of $w=-2\log{\cal L}$. 

We found that the three channels for 
the $a_1$ resonant substructure are necessary and sufficient for a description of the data. 
Our model for the $a_1$ substructure is the same as that of E852, but it
is in contrast with the CLEO analysis of $\tau^- \to \nu_{\tau}\pi^+\pi^0\pi^0$ \cite{cleotau}.
The CLEO fit to the $\pi^+\pi^0\pi^0$ line shape required seven different contributions. 
However, the three analyses agree on the general picture of the $a_1$ substructure, according to which 
the dominant contribution comes from the S-wave $\rho(770)^0\pi$, (over 60\% of the $a_1$ decay rate),
followed by the $\sigma\pi$ ($\sim$15\%) and with a small D-wave $\rho(770)^0\pi$ component.

From the results of the $a_1$ resonant substructure we can measure the ratio between the
D- and S-wave $\rho^0\pi^+$ amplitudes. In the flux-tube-breaking model \cite{isgur} this ratio
is given by 

\begin{equation}
 \frac{ A(a_1 \to (\rho\pi)_D)}{A(a_1 \to (\rho\pi)_S)} = \frac{-D}{\sqrt{32}S},
\end{equation}
where $D$ and $S$ are the D- and S-wave amplitudes. Using this definition we measure

\begin{equation}
\frac{ A(a_1 \to (\rho\pi)_D)}{A(a_1 \to (\rho\pi)_S)} = -0.043 \pm 0.009 \pm 0.005,
\end{equation}
where the statistical uncertainty is obtained from the full covariance matrix.

The $a_1(1260)$ line shape was determined by the iterative procedure described above.
We scanned over the values $s_0^{a_1}$ and $\Gamma_0^{a_1}$ in order to find the
combination of values that optimizes the fit result. The resonant substructure remained stable 
during this scanning procedure; that is, the values of $c_i$ varied less than 10\%. We found that
there is some correlation between $s_0^{a_1}$, $\Gamma_0^{a_1}$ and the coefficients $c_i$. 
Fits of nearly equivalent quality can be
obtained using combinations of these parameters along the diagonal of a rectangle, in the  
$\sqrt{s_0^{a_1}} \times \Gamma_0^{a_1}$ plane, with sides $1230  <\sqrt{s_0^{a_1}}< 1270 ~\mathrm{MeV}/c^2$ and 
$520  < \Gamma_0^{a_1} < 680 \mathrm{~MeV}/c^2$. The optimum values are 
$\sqrt{s_0^{a_1}} = 1240 ~\mathrm{MeV}/c^2$ and $\Gamma_0^{a_1} = 560 ~\mathrm{MeV}/c^2$. 

The total width $\Gamma_{\mathrm{tot}}^{a_1}(s)$ (Eq. 8) is shown in Fig. \ref{gamma-a1}, as a function of the
$\pi^+\pi^-\pi^+$ mass squared. The solid line is the total width, whereas the dashed line represents the
contribution from the $K^*K$ channel. The opening of this channel introduces a cusp-like effect on the
total width.

Our values of the $a_1$ parameters differ significantly from CLEO result:
$\sqrt{s_0^{a_1}} = (1331 \pm 10) ~\mathrm{MeV}/c^2$ and $\Gamma_0^{a_1} = (814 \pm 38) ~\mathrm{MeV}/c^2$. 
The CLEO values, however, depend strongly on their form factor parameter $r$. When the value of this parameter is set
to  $r=1.2 ~\mathrm{GeV}^{-1}$ in the CLEO analysis, their values of $\sqrt{s_0^{a_1}}$ and $\Gamma_0^{a_1}$ are in very good
agreement with the result of this analysis.
Our result is in good agreement with the model of K\"{u}hn and Santamaria \cite{kuhn} 
($\sqrt{s_0^{a_1}} = (1262 \pm 11) ~\mathrm{MeV}/c^2$ and $\Gamma_0^{a_1} = (621 \pm 66) ~\mathrm{MeV}/c^2$) 
and a bit higher than the values from Isgur {\em et al.} 
($\sqrt{s_0^{a_1}} = (1210 \pm 7) ~\mathrm{MeV}/c^2$ and 
$\Gamma_0^{a_1} = (457 \pm 23) ~\mathrm{MeV}/c^2$) \cite{isgur}.
Both models consider only the $\rho\pi$ mode for the $a_1$ resonant substructure.

\begin{figure}
\includegraphics[width=8.5 cm]{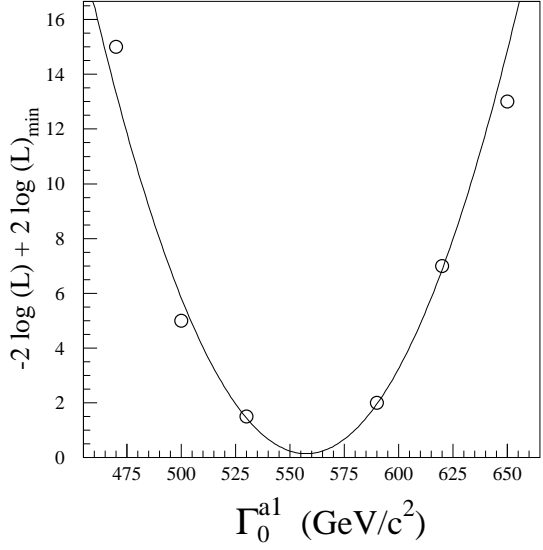}
\caption{The difference $-2\log{\cal L} + 2\log{\cal L}_{\mathrm{max}}$ as a function of the $\Gamma_0^{a_1}$ 
parameter, for $\sqrt{s_0^{a_1}}=$ 1240 MeV/$c^2$.}
\label{a1fcn}
\end{figure}

\begin{figure}
\includegraphics[width=8.5 cm]{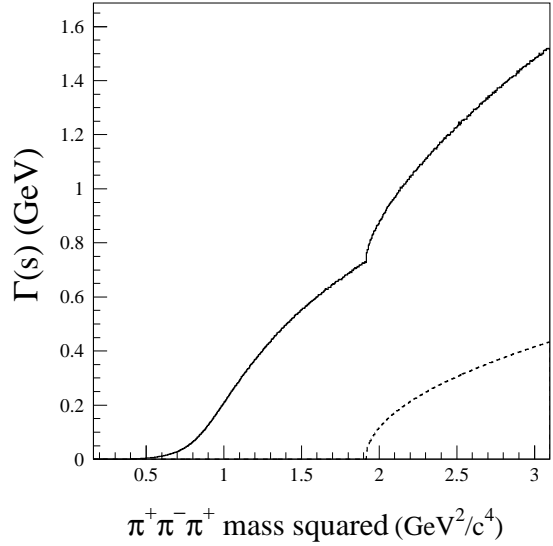}
\caption{The $s$-dependent $a_1$ width, $\Gamma^{a_1}_{\mathrm{tot}}(s)$, plotted as a function of the
$\pi^+\pi^-\pi^+$ mass squared, $s$. The dashed line represents the contribution from $a_1 \to K^*K$.}
\label{gamma-a1}
\end{figure}

\subsection{$D^0 \to \rho^0\rho^0$ results}

The amplitude for the $D \to \rho^0 \rho^0$ decay is written in the helicity formalism. The most general 
amplitude $D \to V_1(\lambda_1) V_2(\lambda_2)$ is a sum of three independent components, corresponding to 
the linear polarization states of the vector mesons. The $D^0$ is an initial state with $J=0$, so the 
vector mesons helicities $\lambda_1$ and $\lambda_2$ must be equal, $\lambda_1=\lambda_2=\lambda=-1,0,+1$. The 
three independent helicity amplitudes, corresponding to each of the possible helicity states are
$H_-$, $H_0$, and $H_+$.
Alternatively, one can decompose the $D \to V_1 V_2$ decay amplitude in the so-called transversity basis
\cite{rosner},  with amplitudes given by $A_0=H_0$, $A_{\parallel}=(H_++H_-)/\sqrt{2}$ and 
$A_{\perp}=(H_+-H_-)/\sqrt{2}$. 
This basis represents states in which the vector mesons are longitudinally polarized ($A_0$), or 
transversely polarized ($A_{\parallel}$, $A_{\perp}$). The transversity basis states are parity eigenstates.

In the $D$ rest frame the two vector mesons are back-to-back. After a boost along the line-of-flight to 
the $V_1$ rest frame, we define the helicity angle $\theta_1$ as the angle between the $\pi^-$ and the 
direction of the boost, measured counterclockwise. The helicity angle $\theta_2$ is defined in a similar 
way. With these definitions and the acoplanarity angle, $\chi$, defined as the angle between the $V_1$, 
$V_2$ decay planes measured in the $D$ frame, one can write the $D\to V_1 V_2$ decay amplitude as

\begin{equation}
{\cal A}(D\to\rho\rho) \propto A_0 \cos \theta_1\cos \theta_2 + 
\frac{A_{\parallel}}{\sqrt{2}}\sin \theta_1 \sin \theta_2 \cos \chi -
i\frac{A_{\perp}}{\sqrt{2}} \sin \theta_1 \sin \theta_2 \sin\chi.
\end{equation}

In the above equation the amplitudes $A_0$, $A_{\parallel}$, and $A_{\perp}$ are of the form 
$A_i = a_ie^{i\delta_i}f(m_{\pi\pi}^2)$, where $a_i$ and $\delta_i$ are real constants and $f(m_{\pi\pi}^2)$ 
has all the energy dependent terms: Breit-Wigner functions, 
form factors, and the energy dependent part of the angular distribution.

The longitudinal polarization  is defined as the ratio of the longitudinal to the total decay rate,

\begin{equation}
P_L = \frac{\left|A_0 \right|^2}{\left|A_0 \right|^2 + \left|A_{\parallel} \right|^2 + \left|A_{\perp} \right|^2 }.
\end{equation}

Using the values obtained by the fit (Table \ref{fractions}) we measured $P_L = (71 \pm 4 \pm 2)\% $.

Our value is in agreement with the {\sc BaBar} result from $B^0 \to \rho^0\rho^0$ \cite{babarvv2}, 
$P_L = (87 \pm 13 \pm 4)\% $, but is somewhat different from the polarization measured in 
$B^0 \to \rho^+\rho^-$ ($P_L = (99 \pm 3 \pm 4)\% $) \cite{babar} and in 
$B^{\pm} \to \rho^{\pm}\rho^0$ ($P_L = (95 \pm 11 \pm 2)\% $ \cite{belle} and
$P_L = (90.5 \pm 4.2 \pm 2.5)\% $) \cite{babarvv}.

The polarization measured in $B$ decays is in agreement with theoretical expectations
\cite{kramer}, which are based on the assumption that the transition amplitude for the $B \to \rho\rho$
can be written as a product of two independent hadronic currents.
The factorization hypothesis may be a good approximation for $B$ meson decays, due to the large value of 
its mass, but fails to describe hadronic decays of $D$ mesons. That could explain the observed 
difference in the values of the longitudinal polarization fraction from $D$ and $B$ decays.

\subsection{Systematic uncertainties}
\label{sys}

The representation of the structures present in  $\pi^-\pi^+\pi^-\pi^+$ is a major
source of systematic uncertainty. The problem starts with the choice of a set of amplitudes to describe 
the data. Furthermore, our model depends on some input parameters, which are fixed during the fit: the parameter $r_R$
for the form factors; the $a_1$ parameters $s_0^{a_1}$, $\Gamma_0^{a_1}$ and $g_{K^*K}$; the relative 
amount of the backgrounds. In addition, there are the parameters defining the $\sigma$ and the $f_0(980)$ line shape:
$m_0^{\sigma}$ and $\Gamma_0^{\sigma}$, and $g_{\pi}$, $g_K$, and $m_0^{f_0}$. 
We should also mention that there is some freedom in the choice of the angular distributions.

We investigated the effect of the uncertainty on the input parameters of our model on the fractions and
phases for the various intermediate channels. For this purpose, 
the data was fit with variations of our baseline model. We varied the value of $r_R$ from 1 to 5 
$\mathrm{GeV}^{-1}$. We have also fit the data using different sets of $a_1$ parameters,
$s_0^{a_1}$, $\Gamma_0^{a_1}$ and $g_{K^*K}$. For the $\sigma$ and 
$f_0(980)$ line shapes we used the BES parameterization \cite{bes2,bes3}. 

In addition, we varied the background parameterization: the
relative amount of background, the background composition, adding a possible contribution from
uncorrelated $\rho^0\rho^0$ events, and the relative fractions of the backgrounds, changing the 
fraction of $\rho^0\pi^+\pi^-$ from 0 to 50\%.

A very small change of the values of the fit parameters, compared to their central values, is observed 
when we vary the $r_R$ parameter, the background parameterization and the $f_0(980)$ line shape. 
The systematic errors are largely dominated by the uncertainty in the $a_1$ parameters and on the 
$\sigma$ line shape, in this order. The total systematic errors are taken as the root mean square of 
fit variations. These are propagated to fractions, the longitudinal polarization measurement, and the D/S ratio. 
The systematic errors are quoted in Table \ref{fractions}.

\section{Conclusions}

In this paper we presented a new measurement of the relative branching ratio
$\Gamma(D^0 \to\pi^+\pi^- \pi^+\pi^-)/\Gamma(D^0 \to K^-\pi^+\pi^-\pi^+) = 0.0914 \pm 0.0018 \pm 0.0022$,
based on a very clean sample of 6360$\pm$115 events.
Our value is compatible with the recent CLEO-c result \cite{CLEOc}, which has a sample of comparable size, but with
a higher background level.

A full, coherent amplitude analysis of the $D^0 \to \pi^-\pi^+\pi^-\pi^+$ decay was made for the first time.
The isobar model with nine resonant amplitudes provides a description of the general features of the data.
In our model there are contributions of three types. The dominant contribution comes from
the decay $D^0 \to a_1(1260)^+\pi^-$, accounting for  60\% of the total decay rate. This is followed
the decay $D^0 \to \rho(770)^0\rho(770)^0$, whose relative fraction amounts to 25\%, and by
the three-body nonresonant decays $D^0 \to \sigma\pi^+\pi^-$, $D^0 \to f_0(980)\pi^+\pi^-$ and 
$D^0 \to f_2(1270)\pi^+\pi^-$, with a combined fraction of 20\%. 

The $D^0 \to \pi^-\pi^+\pi^-\pi^+$ decay is a very suitable tool for investigation of the $a_1(1260)$ meson.
A good description of this state is achieved assuming a simple resonant substructure, with only three
contributions: a dominant $a_1^+ \to \rho(770)^0\pi^+$ S-wave, a small $a_1^+ \to \rho(770)^0\pi^+$ D-wave, and the 
$a_1^+ \to \sigma \pi^+$. In the framework of the flux-tube-breaking model, we measure the $a_1^+ \to \rho(770)^0\pi^+$
D/S ratio to be $R=-0.043 \pm 0.009 \pm 0.005$.

The line shape of the $a_1(1260)$ is also an output of the amplitude analysis. The best description of
our data is achieved with the $a_1(1260)$ Breit-Wigner parameters, $\sqrt{s_0^{a_1}}$ and $\Gamma_0^{a_1}$,
in the range $1230  <\sqrt{s_0^{a_1}}< 1270 ~\mathrm{MeV}/c^2$ and $520  < \Gamma_0^{a_1} < 680 ~\mathrm{MeV}/c^2$,
with optimum values at $\sqrt{s_0^{a_1}} = 1240 ~\mathrm{MeV}/c^2$ and $\Gamma_0^{a_1} = 560 ~\mathrm{MeV}/c^2$.

The $D^0 \to \pi^-\pi^+\pi^-\pi^+$ decay has an important $\rho^0\rho^0$ component. The  helicity formalism
is used to to describe the angular distribution of the four pions from this mode. Using the transversity
basis states and a full amplitude analysis, we measure the ratio of the longitudinal polarization to the total 
$D \to \rho^0\rho^0$ rate to be $P_L=(71 \pm 4 \pm 2)$\%. 

The fit quality, from an inspection of the five-dimensional phase space, is poor and it is not improved 
by changing the representation of individual signal amplitudes or by adding other amplitudes to the baseline model.
We conclude that in the case of  $D^0 \to \pi^-\pi^+\pi^-\pi^+$ the failure of the isobar model in providing 
an accurate description of our data suggests that other type of effects should be taken into account.
Possible candidates are the Bose-Einstein correlations or a non-uniform nonresonant amplitude.
The most crucial missing ingredient may be the energy dependent final state interactions. 
Unfortunately this is a very difficult problem, 
even for three-body decays. To our knowledge, there are no calculations of the effects of FSI in four-body decays. 
This problem, however, must be addressed in order to explore the full potential of the hadronic decays of heavy flavor and
the amplitude analysis technique.

\begin{acknowledgments}
We wish to acknowledge the assistance of the staffs of Fermi National
Accelerator Laboratory, the INFN of Italy, and the physics departments of
the collaborating institutions. This research was supported in part by the
U.~S. National Science Foundation, the U.~S. Department of Energy, the
Italian Istituto Nazionale di Fisica Nucleare and Ministero della Istruzione
Universit\`a e Ricerca, the Brazilian Conselho Nacional de Desenvolvimento
Cient\'{\i}fico e Tecnol\'ogico, CONACyT-M\'exico, and the Korea Research
Foundation of the Korean Ministry of Education.
\end{acknowledgments}

\appendix

\section{Angular distributions}

The angular distributions of the final state particles result from 
angular momentum conservation throughout the whole decay chain. These
distributions depend on both the spin of the intermediate resonances
and on the orbital angular momentum at each vertex. In the case of
sequential decays, like the $D \to a_1^+ \pi^-$ , $a^+_1 \to \rho^0 \pi^+$, 
$\rho^0 \to \pi^+ \pi^-$ chain, parity conservation is also a constraint on the $a_1$ decay and all 
its subsequent decay products. 

Amplitudes describing angular distributions, hereafter called spin amplitudes
for short, are, in general defined using the relativistic tensor formalism \cite{filippini,chung}.
This formalism explores the connection between the only available observables ---
the momenta of the final state particles --- and the spin/orbital angular momentum 
dynamics. Polarization vectors, representing the spin/orbital angular momentum 
involved at each vertex, as well as all relevant 4-momenta, are written in terms of 
these final state momenta.  Covariance, invariance under rotations and parity 
conservation, whenever applicable, restricts the many possible combinations between 
4-momenta and polarization vectors.

The relativistic tensor formalism was used in this analysis to describe
all intermediate channels, except for the $D^0 \to \rho^0\rho^0$, for which we
used the helicity formalism.

\subsection{$D \to a_1\pi$}

In our model for the $a_1$ resonant substructure there are three amplitudes: 
$a_1^+ \to \rho^0\pi^+$, with the $\rho^0$ and the $\pi^+$ in relative S- and D-wave;
and the $a_1^+ \to \sigma\pi^+$. We denote the  $a_1$ and $\rho^0$ polarization 4-vectors
with helicity $\lambda$ by $\varepsilon_{\mu}(\lambda)$ and $e_{\mu}(\lambda)$, respectively.

We form Lorentz scalars at each step of the decay sequence by contracting these polarization 
vectors with the appropriate 4-momenta. In a step where the resonance appears as a decay product, 
the complex conjugate of its polarization 4-vector is used. If the $a_1$ is formed by 
particles 1, 2 and 3, then the amplitude for the $D^0 \to a_1^+ \pi_4^-$ is

\begin{equation}
\langle a_1\pi_4 \left|A \right|D\rangle \propto \varepsilon^*_{\mu}(\lambda)p_4^{\mu}.
\end{equation}

The amplitude for the $a_1^+ \to \rho^0 \pi^+$ ($\rho^0 \pi^+$ in S-wave) decay is

\begin{equation}
\langle \rho \pi \left|A \right|a_1\rangle \propto \varepsilon_{\nu}(\lambda)e^{*\nu}(\lambda '),
\end{equation}
and ($\rho^0 \pi^+$ in  D-wave)

\begin{equation}
\langle \rho \pi \left|A \right|a_1\rangle \propto \varepsilon_{\nu}(\lambda) 
Q^{\nu} e^*_{\sigma}(\lambda')p_{a_1}^{\sigma},
\end{equation}
with $Q=p_{\rho}-p_3$,  ~$p_{\rho}$, $p_{\sigma} = p_1+p_2$ and $p_{a_1} = p_1+p_2+p_3$

The amplitude for the $\rho^0$ decay is

\begin{equation}
\langle \pi \pi \left|A \right|\rho\rangle \propto e_{\alpha}(\lambda ') q^{\alpha},
\end{equation}
with $q=p_1-p_2$. Finally, the amplitude for the $a_1^+ \to \sigma \pi^+$ is

\begin{equation}
\langle \sigma \pi \left|A \right|a_1\rangle \propto \varepsilon_{\nu}(\lambda)Q^{\nu}.
\end{equation}

We then combine the amplitudes for each step and must sum over the $a_1$ and $\rho^0$ polarizations, since 
none of these states is observed directly. The sum over polarizations is simplified by using the 
completeness relation \cite{chung}

\begin{equation}
\sum_{\lambda} = \varepsilon^*_{\mu}(\lambda)\varepsilon_{\nu}(\lambda) = -g_{\mu\nu} + p_{\mu}p_{\nu}/p^2.
\end{equation}

After summing over the unobserved polarizations, the spin amplitude for 
$D^0 \to a_1^+ \pi_4^-$, $a_1^+ \to \rho^0 \pi^+$ S-wave, $\rho^0 \to \pi^+\pi^-$ is

\begin{equation}
{\cal M} = p_4^{\mu}\left(-g_{\mu\nu} + \frac{p_{\mu}^{a_1}p_{\nu}^{a_1}}{p_{a_1}^2}\right)
\left(g^{\nu\sigma} - \frac{p^{\nu}_{\rho}p^{\sigma}_{\rho}}{p_{\rho}^2}\right)q_{\sigma}.
\end{equation}

For the $a_1^+ \to \rho^0 \pi^+$ D-wave the spin amplitude reads

\begin{equation}
{\cal M} = p_4^{\mu}\left(-g_{\mu\nu} + \frac{p_{\mu}^{a_1}p_{\nu}^{a_1}}{p_{a_1}^2}\right)
Q_{\nu}p_{a_1}^{\sigma}
\left(-g_{\sigma\alpha} + \frac{p_{\sigma}^{\rho}p_{\alpha}^{\rho}}{p_{\rho}^2}\right)q^{\alpha}.
\end{equation}

Finally, the spin amplitude for the $a_1^+ \to \sigma \pi^+$ is

\begin{equation}
{\cal M} = p_4^{\mu}\left(-g_{\mu\nu} + \frac{p_{\mu}^{a_1}p_{\nu}^{a_1}}{p_{a_1}^2}\right)Q^{\nu}.
\end{equation}

\subsection{$D \to \rho^0\rho^0$}

The spin amplitudes for the $D \to \rho^0\rho^0$ decay are written in the helicity formalism. This decay
is of the type $P\to V_1(\lambda_1)V_2(\lambda_2)$, where $\lambda_1,\lambda_2$ are the vector
meson helicities. Given that the initial state has $J=M=0$, the constraint 
$\left|\lambda_1 - \lambda_2 \right|\leq M$ implies that $\lambda_1 = \lambda_2 = \lambda$.
The only allowed values for the vector meson helicity are $\lambda = -1$, $0$, and $+1$. The helicity-basis
states $|JM,\lambda\rangle $ are $|00,-1\rangle $, $|00,0\rangle $ and $|00,+1\rangle$.

For each helicity state there is one independent amplitude. These amplitudes are functions of the helicity
angles, defined in each resonance rest frame, and of the angle formed by the vector mesons decay planes
(acoplanarity angle), measured in the $D$ rest frame. 
We define the $z$ axis along the line of flight of
the vector mesons in the $D$ rest frame. With the notation $\rho_1^0 \to \pi^-_1\pi^+_2$ and 
$\rho_2^0 \to \pi^-_3\pi^+_4$, the helicity angles $\theta_1$ and $\theta_2$ are defined as the angle
between $\pi_1$ and the $z$ axis, in the $\rho_1^0$ rest frame, and the angle between $\pi_3$ and the $z$ axis, 
in the $\rho_2^0$ rest frame. 

Now with all momenta measured in the $D$ frame, the orientations
of the decay planes are $\hat{n}_1 = \vec{p}(\pi^+_2) \times \vec{p}(\pi^-_1)$ and 
$\hat{n}_2 = \vec{p}(\pi^+_4) \times \vec{p}(\pi^-_3)$. The acoplanarity angle is 
$\chi=\hat{n}_1 \cdot \hat{n}_2$.

With the above definitions of $\theta_1, ~\theta_2$ and $\chi$, the general form of the
helicity amplitudes is

\begin{equation}
A_{\lambda} = e^{i\lambda\chi} d^1_{\lambda,0} (\theta_1) d^1_{\lambda,0} (\theta_2).
\end{equation}

With the appropriate $d$ functions we have

\begin{equation}
A_{-1} = \frac{1}{2} e^{-i\chi} \sin{\theta_1} \sin{\theta_2},
\end{equation}

\begin{equation}
A_{0} = \cos{\theta_1}\cos{\theta_2},
\end{equation}

\begin{equation}
A_{+1} = \frac{1}{2} e^{i\chi} \sin{\theta_1} \sin{\theta_2}.
\end{equation}

The helicity-basis states are not parity eigenstates. One can construct a new basis with definite parity,

\begin{equation}
|f_{\parallel}> = \frac{|00,1\rangle ~+ ~|00,-1\rangle}{\sqrt{2}},
\end{equation}

\begin{equation}
|f_{\perp}> = \frac{|00,1\rangle ~- ~|00,-1\rangle}{\sqrt{2}},
\end{equation}

\begin{equation}
|f_L> = |00,0\rangle .
\end{equation}

This is known as the transversity basis \cite{rosner}. The corresponding amplitudes are

\begin{equation}
A_{\parallel} = \frac{A_{+1} ~+ ~A_{-1}}{\sqrt{2}} = 
\frac{1}{\sqrt{2}} \cos{\chi} \sin{\theta_1} \sin{\theta_2},
\end{equation}

\begin{equation}
A_{\perp} = \frac{A_{+1} ~- ~A_{-1}}{\sqrt{2}} =
\frac{i}{\sqrt{2}} \sin{\chi} \sin{\theta_1} \sin{\theta_2},
\end{equation}

\begin{equation}
A_L = A_0 = \cos{\theta_1}\cos{\theta_2}.
\end{equation}

The above amplitudes are expressed in terms of non-covariant quantities, defined in different 
reference frames. Covariance is recovered by multiplying the helicity amplitudes by an energy dependent 
term, which accounts for the boosts from each resonance frame to the $D$ frame. Each of the above
amplitudes is then multiplied by

\begin{equation}
A_{\lambda} \to A_{\lambda} \times \sqrt{\frac{\Gamma(s)\sqrt{s}}{p^*(s)}},
\end{equation}
where $s$ is the two-pion invariant mass, $p^*(s)$ is the break-up momentum and $\Gamma(s)$ is

\begin{equation}
\Gamma(s) = \Gamma_0 \frac{\sqrt{s_0}}{\sqrt{s}} \left( \frac{p^*(s)}{p^*(s_0)} \right)^3
\frac{1+(r_Rp^*(s_0))^2}{1+(r_Rp^*(s))^2}.
\end{equation}

These are the formulae used for the $D \to \rho^0\rho^0$ spin amplitudes.

\newpage %Just because of unusual number of tables stacked at end

\end{document}